\documentclass[twocolumn,notitlepage,prl,superscriptaddress,showpacs]{revtex4-1}
\usepackage{hyperref}
\usepackage[usenames,dvipsnames]{color}
\usepackage{amsmath}
\usepackage[english]{babel}
\usepackage{amssymb}
\usepackage{graphicx}
\usepackage{epstopdf}
\begin{document}

\title{Knight shifts, nuclear spin-relaxation rates, and spin echo decay times in the pseudogap regime of the cuprates: Simulation and relation to experiment}

\author{Xi Chen}
\affiliation{Department of Physics, University of Michigan, Ann Arbor, Michigan 48109, USA} 
\author{J.P.F. LeBlanc}
\affiliation{Department of Physics, University of Michigan, Ann Arbor, Michigan 48109, USA} 
\author{Emanuel Gull}
\affiliation{Department of Physics, University of Michigan, Ann Arbor, Michigan 48109, USA}

\date{\today}

\begin{abstract}
We study the temperature and doping evolution of the NMR Knight shift, spin relaxation rate, and spin echo decay time in the pseudogap regime of the two-dimensional Hubbard model for parameters believed to be relevant to cuprate superconductors using cluster dynamical mean field theory. 
We recover the suppression of the Knight shift seen in experiment upon entering the pseudogap regime and find agreement between single and two-particle measures of the pseudogap onset temperature $T^*$. 
The simulated spin-echo decay time shows a linear in $T$ behavior at high $T$ which flattens off as $T$ is lowered, and increases as doping is increased. 
The relaxation rate shows a marked increase as $T$ is lowered but no indication of a pseudogap on the Cu site, and a clear downturn on the O site, consistent with experimental results on single layer materials but different from double layer materials. 
The consistency of the simulated susceptibilities with experiment, along with similar agreement on the single-particle level and the absence of long-range order and symmetry breaking suggests that the pseudogap  is well described by strong short-range correlation effects and that long-range order and multi-orbital effects are not required.
\end{abstract}

\maketitle
The pseudogap in the cuprates was discovered as a reduction of the knight shift ($K_s$) and spin relaxation time $T_1$ measured in nuclear magnetic resonance (NMR) experiments \cite{Warren89,alloul:1989,Walstedt90,Takigawa1991,RussBook}.
Subsequent experimental research \cite{Timusk99,Hufner08} has resulted in its detection in a wide range of materials and experimental probes for dopings smaller than optimal doping and temperatures smaller than 300$K$. In single particle experiments, the pseudogap shows as a clear suppression of the density of states near the antinodal, but not the nodal, point of the Brillouin zone and is well described by non-perturbative numerical simulations  \cite{Huscroft01,Tremblay06,Macridin06,Kyung06,Werner098site,Gull09_8site,Gull10_clustercompare,Sordi12} of fermion model systems.

 Unlike single-particle probes, NMR provides a direct measure of a two-particle quantity, the magnetic (spin) susceptibility.  The complete theoretical understanding of the two-particle signals measured in NMR is difficult, requiring two components: a precise relation of the NMR signal to correlation functions \cite{Mila1989,Shastry1989,Barzykin1995} and the low-energy spin susceptibility, and a reliable calculation of the spin susceptibility itself. While the first aspect has been well understood, directly obtaining the spin susceptibility of a correlated system has proven to be a formidable task, and as a result theoretical calculations of the NMR response have been limited to analytic or semi-analytic methods \cite{Bulut1990,Bulut92}, high temperature \cite{Scalapino93}, or sign-problem free attractive models \cite{Randeria92}.  While these calculations provide a qualitative understanding of NMR signals outside the strong correlation regime, they do not contain a pseudogap in the single particle quantities and therefore cannot comment on the role of a pseudogap in NMR spectra.
 
Recent advances in the numerical simulation of interacting fermionic lattice models \cite{LeBlanc15} have made simulation of susceptibilities possible. In particular, a combination of cluster dynamical mean field methods \cite{Maier05} with continuous-time \cite{Gull11_RMP} auxiliary field \cite{Gull08_ctaux,Gull11_submatrix} impurity solver extensions to two-particle functions \cite{Lin12} now allow for the unbiased calculation of generalized susceptibilities \cite{Chen15}.

 In this paper, we show elements of the magnetic susceptibility that are directly related to NMR experiments: the Knight shift, the spin echo decay rate, and the relaxation rate. We focus on the temperature and doping dependence of these quantities for which a large body of experimental NMR work exists.
 Our results show remarkable similarity  (both in temperature and doping dependence) to the experimentally measured quantities, indicating that the single-orbital Hubbard model, away from half filling and with an interaction strength close to the bandwidth captures much of the two-particle physics observed in experiment.  Further, we isolate the role of pseudogap physics in each NMR probe.

We study the single orbital Hubbard model in two dimension with nearest and next nearest hopping parameters in the normal state,
\begin{align}
H = \sum_{k,\sigma} \left(\epsilon_{k} -\mu\right)c_{k\sigma}^\dagger c_{k\sigma}+U\sum_i n_{i\uparrow}n_{i\downarrow},
\label{eq:H}
\end{align}
where $\mu$ is the chemical potential, $k$ momentum, $i$ labels of sites in real-space, $U$ the interaction, and the dispersion is given by
$\epsilon_k=-2t\left[\cos(k_x)+\cos(k_y)\right]-4t'\cos(k_x)\cos(k_y).$

The generalized susceptibility $\chi$ \cite{rohringer} is written in imaginary time in terms of the one- and particle $G_{\sigma_1\sigma_2}(k_1\tau_1,k_2\tau_2)$ $=$ $\langle T_\tau(c^\dagger_{k_1\sigma}(\tau_1)c_{k_2\sigma}(\tau_2)\rangle$ and two-particle $G_{2,\sigma_1\sigma_2\sigma_3\sigma_4}(k_1\tau_1,k_2\tau_2,k_3\tau_3,k_4\tau_4)$ $=$ $\langle T_\tau(c^\dagger_{k_1\sigma}(\tau_1)c_{k_2\sigma}(\tau_2)c^\dagger_{k_3\sigma}(\tau_3)c_{k_4\sigma}(\tau_4)\rangle$ Green's functions as
$\chi_{\sigma_1\sigma_2\sigma_3\sigma_4}(k_1\tau_1,k_2\tau_2,k_3\tau_3,k_4\tau_4)$ $=$ $G_{2,\sigma_1\sigma_2\sigma_3\sigma_4}(k_1\tau_1,k_2\tau_2,k_3\tau_3,k_4\tau_4)$ $-$ $G_{\sigma_1\sigma_2}(k_1\tau_1,k_2\tau_2)$ $G_{\sigma_3\sigma_4}(k_3\tau_3,k_4\tau_4)$.
Its Fourier transform is
\begin{align}
\nonumber&\chi_{ph\sigma\sigma'}^{\omega\omega'\nu}(k,k',q) = \int_{0}^\beta d\tau_1d\tau_2d\tau_3 e^{-i\omega\tau_1+i(\omega+\nu)\tau_2-i(\omega'+\nu)\tau_3}
\\ &\times \chi_{\sigma\sigma\sigma'\sigma'}(k\tau_1,(k'+q)\tau_2,(k+q)\tau_3,k'0)
\end{align}
where $\omega$ and $\omega'$ are fermionic Matsubara frequencies, $\nu$ is a bosonic Matsubara frequency, $\sigma$ and $\sigma'$ are $\uparrow$ or $\downarrow$ spin labels and $k$, $k'$ and $q$ are initial, final and transfer momenta respectively. $ph$ denotes the Fourier transform convention \cite{rohringer}.  The main object of interest, the spin susceptibility, is then defined as
\begin{align}
\chi_m = \chi_{ph\uparrow\uparrow}-\chi_{ph\uparrow\downarrow}.
\end{align}

We are interested in three aspects of NMR spectroscopy: the Knight shift $K^S$, the spin echo decay rate $T_{2G}^{-1}$ and the spin lattice relaxation rate $T_1^{-1}$. According to the Mila-Rice-Shastry model \cite{Mila1989,Shastry1989} for hyperfine coupling with itinerant Cu$^{2+}$ holes in high $T_c$ cuprates, the Knight shift $K^S$ measured in nuclear magnetic resonance experiment is proportional to the uniform spin susceptibility,
\begin{align}
K^S\propto\chi_m(q=(0,0),\nu=0).
\label{eq:knight_shift}
\end{align}

For the $^{63}$Cu nuclear spin echo decay rate $^{63}1/T_{2G}$ in paramagnetic state of high $T_c$ cuprates, \textcite{Pennington1991} showed that
\begin{align}
\nonumber^{63}T_{2G}^{-2}=\frac{0.69}{128\hbar^2} \Big[ \frac{1}{N}\sum_q\ {}^{63}F_{\text{eff}}(q)^2\chi'_m(q,0)^2\\
 - \big( \frac{1}{N}\sum_q\ {}^{63}F_{\text{eff}}(q) \chi'_m(q,0) \big)^2 \Big],
\label{eq:T2G}
\end{align}
where $\chi'_m(q,\nu=0)$ denotes the real part of the real-frequency dynamical spin susceptibility at momentum $q$ and frequency $\nu=0$. The prefactor 0.69 originates from the natural abundance of $^{63}$Cu \cite{Imai1993}, and $^{63}F_{\text{eff}}$ is defined in Ref.~\cite{Barzykin1995} with hyperfine coupling constants $A$ and $B$ as $^{63}F_{\text{eff}}=\{A_\parallel+2B[\cos(q_xa)+\cos(q_ya)]\}^2$, $A_\parallel=-4B$.  For simplicity we set $B\equiv 1$ and consider only proportionality.
With this, both $K^S$ and $T_{2G}$ can be calculated directly from a susceptibility on the Matsubara axis since $\chi(q,\nu=0)=\chi(q,i\nu=0)$. 

The spin-lattice relaxation rate $1/T_1$ is related to the imaginary part of dynamical spin susceptibility on the real axis,
\begin{align}
\frac{1}{T_1T} \propto \lim_{\nu\rightarrow0}\sum_q ~^{\alpha}F_{\parallel}(q)\frac{\chi''_m(q,\nu)}{\nu}.\label{inverseT1}
\end{align}
where $^{\alpha}F_{\parallel}(q)$ differs for $^{63}Cu$ and $^{17}O$, as defined in Ref.~\cite{Barzykin1995}. 
\begin{align}
\nonumber &^{63}F_{\parallel}=\{A_{\perp}+2B[\cos(q_x)+\cos(q_y)]\}^2\\
\nonumber &^{17}F_{\parallel}=2C_{\parallel}^2[1+0.5[\cos(q_x)+\cos(q_y)]]\\
&A_{\perp}=0.84B,  C_{\parallel}=0.91B.
\end{align}
The calculation of $\chi''_m(q,\nu)/\nu$ within a Matsubara formalism requires analytical continuation \cite{Jarrell96}. However, the quantity $S(q,\tau)$, defined as the real-to-k-space Fourier transform of $S_z=n_{i\uparrow}-n_{i\downarrow}$, satisfies
\begin{align}
\nonumber &\sum_q {}^{\alpha}F_{\parallel}(q)S(q,\tau=\frac{1}{2T}) =\sum_q ~^{\alpha}F_{\parallel}(q)\int d\nu \frac{\chi_m''(q,0)}{\nu}\frac{\nu}{\sinh\frac{\nu}{2T}}
\end{align}
(see  Ref.~\cite{Randeria92} for details) such that in the limit $T\rightarrow 0$
\newcommand\limTeq{\overset{T\rightarrow 0}{\propto}}
\begin{align}
\frac{1}{T_1}\limTeq\sum_{kk'q\omega\omega'\nu}\frac{2}{\pi^2N\beta^2}~^{\alpha}F_{\parallel}(q)\chi_m^{\omega\omega'\nu}(k,k',q)e^{\frac{-i\nu}{2T}}.
\label{eq:T1}
\end{align}

The direct numerical solution of Eq.~\ref{eq:knight_shift}, \ref{eq:T2G} and \ref{eq:T1} for the Hubbard Hamiltonian Eq.~\ref{eq:H} is intractable. We therefore employ the dynamical cluster approximation (DCA) \cite{Maier05} which approximates the momentum dependence of the many-body self-energy and irreducible vertex functions by an approximated function that is constant on a set of $N_c$ `patches' in momentum space \cite{Fuhrmann07,Maier05}. The method is a non-perturbative short correlation length approximation and is controlled in the sense that as $N_c$ is increased it converges to the exact limit \cite{Gull11_submatrix,LeBlanc13,LeBlanc15}. 
Throughout this paper we use $N_c$ $=$ $8$, a compromise between accuracy and efficiency that has previously been shown to capture much of the single- \cite{Werner098site,Gull09_8site,Lin10,Gull10_clustercompare,Gull13_raman} and two-particle \cite{Lin12,Chen15} physics observed in experiment and shows a qualitatively correct phase diagram for the pseudogapped and superconducting phase \cite{Gull13_super,Gull12_energy,Gull14_pairing,Gull15_qp}. The interaction strength $U=6t$ is large enough to exhibit a clear pseudogap state but presumably slightly smaller than seen in experiment,  having an optimal doping and pseudogap onset closer to half filling \cite{Chen15}.
We use a next-nearest neighbor hopping of $t'=-0.1t$, and do not allow for long ranged ordered antiferromagnetic or superconducting states.

%All of our results are obtained in the normal state at non-zero temperature in the absence of long-range superconducting or antiferromagnetic order.  Full details are provided in the supplemental material.

\begin{figure}[tbh]
\includegraphics[width=80mm]{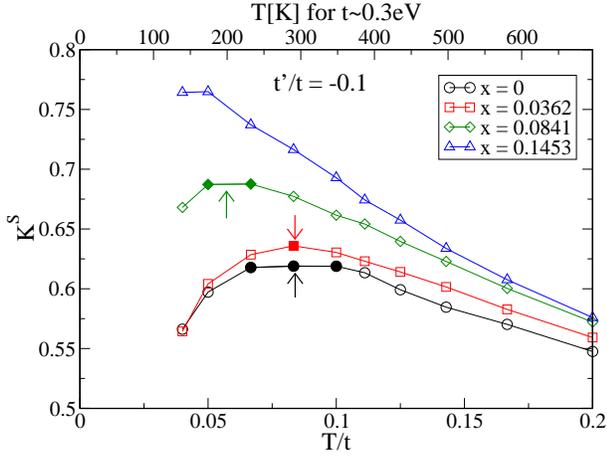}
\caption{Knight shift $K^S\propto\chi_m$ as a function of temperature $T/t$ (lower x-axis) for a series of doping levels computed at $U=6t$, $t'=-0.1t$ obtained from 8-site DCA. Filled symbols: the peak positions of the Knight shift. Arrows: onset of normal state pseudogap obtained by analytical continuation of the single particle spectral function at $K=(0,\pi)$. Upper x-axis: $T/t$ in units of Kelvin assuming $t\sim 0.3eV$.}
\label{fig:knight_shift}
\end{figure}

%\section{discussion of figures}
In Fig.~\ref{fig:knight_shift} we present the simulated NMR Knight shift, Eq.~\ref{eq:knight_shift}, as a function of temperature $T/t$ for a several dopings $x$. For these parameters, the largest $T_c$ on the hole doped side is $T_c=0.03t$ at $x=0.09$. At large doping $x=0.1453$ (triangle, solid blue line), the simulated Knight shift monotonically increases as $T$ is reduced.  Doped cases show a maximum at a temperature $T^*_{Ks}$, indicated by filled symbols. As the density decreases from $x=0.0841$ to $x=0$, this $T^*_{Ks}$ gradually moves to higher $T$. At all temperatures studied, the overall magnitude of the Knight shift increases as doping is increased.

Several features in Fig.~\ref{fig:knight_shift} described above are consistent with what is observed in NMR experiment on high $T_c$ cuprates. Firstly, in the underdoped regime the downturn of $\chi$ as $T$ is lowered is widely observed in $K_s$ at various nuclei sites, see {\it e.g.} Fig.~8 in Ref.~\cite{Takigawa1991} on YBa$_2$Cu$_3$O$_{6.63}$ and Fig.~7 in Ref.~\cite{Curro1997} on YBa$_2$Cu$_4$O$_8$; and similar data for $\chi(T)$ is found in squid magnetometry  of La$_{2-x}$Sr$_x$CuO$_4$ \cite{Nakano1994}, which has historically been interpreted as the onset of the pseudogap phase \cite{Takigawa1991}. Secondly, the increasing Knight shift with increasing doping is observed in a wide range of compounds, including La$_{2-x}$Sr$_x$CuO$_4$ \cite{Nakano1994,Ohsugi1994}, YBa$_2$CuO$_{7-x}$ and YBa$_2$Cu$_4$O$_8$ \cite{Barzykin1995}, and Y$_{1-x}$Pr$_x$Ba$_2$Cu$_3$O$_7$ \cite{Barzykin1995}.

At high temperature, there is a distinct difference between the susceptibility measured in the bilayer material YBa$_2$CuO$_{6.63}$, which displays a broad maximum at $K=500$ and remains approximately constant up to 630K \cite{Curro1997}, and that of the single layer material La$_{2-x}$Sr$_x$CuO$_4$ \cite{Johnston1989,Nakano1994}, where measurements indicate a slowly decreasing knight shift above $T^*$. This discrepancy may be caused by magnetic coupling of copper-oxygen planes in the bilayer materials. Our calculations, which are done on a purely two-dimensional system, are consistent with measurements performed on single layer materials.

The arrows in Fig.~\ref{fig:knight_shift} indicate the onset temperature of the pseudogap in the single particle spectral function calculated by analytical continuation of the single particle Green's function \cite{Lin10} using the maximum entropy method \cite{Jarrell96,Levy2016}. From the temperature evolution of $A_{K=(\pi,0)}(\omega)$, we define $T^*$ as the temperature at which a suppression of the density of states appears near zero frequency (see Fig.~S1 in supplemental material). In agreement with Ref.~\cite{Vidhyadhiraja09,Sordi13}, $T^*_{Ks}$ exhibits the same dependence on temperature and doping level as $T^*$, showing crossover temperatures identified with single-particle quantities (density of states) and two-particle quantities (Knight shift) to be the same \footnote{While this work uses lattice susceptibility to calculate the Knight shift, Ref.~\cite{Vidhyadhiraja09} identifies $T^*$ based on cluster susceptibility, whose doping dependence is not consistent with NMR experiment.}.

\begin{figure}[tbh]
\includegraphics[width=80mm]{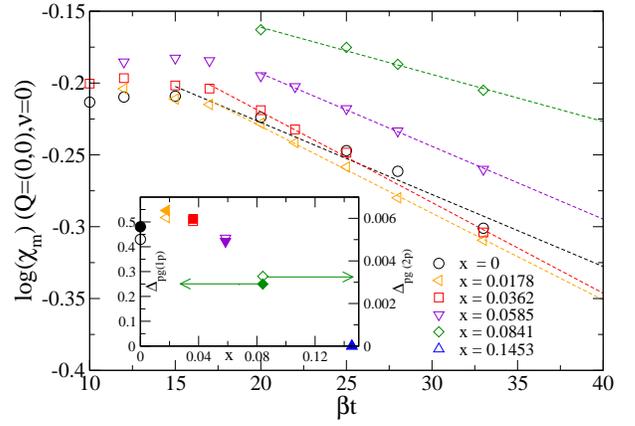}
\caption{Extraction of $\Delta_{pg(2p)}$ from Knight shift data via $\chi_m(T)=\chi_0 \exp(-\Delta_{pg(2p)}/T)$. Open symbols: data of Fig.~\ref{fig:knight_shift} plotted as $log(\chi_m)$ vs. $\beta$. Dashed lines: linear fits to the data in exponentially decaying regime. Inset: comparison between pseudogap energy extracted from the slope of Arrhenius plot (open symbols, right y-axis) and from the single particle spectral function at $K=(0,\pi)$ (filled symbols, left y-axis).}
\label{fig:Arrhenius}
\end{figure}

Fig.~\ref{fig:Arrhenius} expands further upon the data in Fig.~\ref{fig:knight_shift}, including additional doping levels at $x=0.0178$ and $x=0.0585$ for temperatures above the superconducting $T_c$ and below $T^*_{Ks}$ as an Arrhenius plot. Once a gap has opened, the resulting curves become straight lines within uncertainties, allowing us to extract an energy scale from the slopes using $\chi_m(T)=\chi_0 \exp(-\Delta_{pg(2p)}/T)$. 
The inset of Fig.~\ref{fig:Arrhenius} shows the comparison between the pseudogap energy determined by this method (open symbols) and the corresponding pseudogap energy extracted from the peak-to-peak distance of the single particle spectral function at the antinode (filled symbols). The two energy gaps are proportional as a function of doping.  
The distinct energy scales are however expected since $\Delta_{pg(2p)}$ averages over the Brillouin zone while $\Delta_{pg(1p)}$ only considers the antinodal momenta.  As a result, their actual gap values in this case differ by a factor of 75, independent of doping.  
Similar comparisons for experimental data yields at $x=0.5$ values of $\Delta_{pg(1p)}\approx 150$meV and $\Delta_{pg(2p)}=7.75$meV, a factor of 20 difference \cite{Hufner08,RussBook}.  
If one could obtain a quantitative comparison of this ratio to experiment it might allow for a more precise fit of model parameters than considering single-particle properties alone.

\begin{figure}[tbh]
\includegraphics[width=80mm]{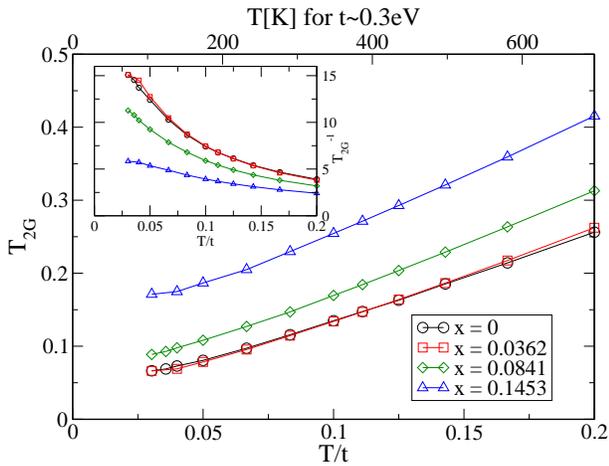}
\caption{Spin echo decay time $T_{2G}$ as a function of temperature for doping level ranging from $x=0$ to $x=0.145$, calculated at $U=6t$, $t'=-0.1t$. Inset: spin echo decay rate $T_{2G}^{-1}$.}
\label{fig:T2G}
\end{figure}

Fig.~\ref{fig:T2G} shows the spin echo decay time $T_{2G}$, a measure of indirect spin-spin coupling, calculated according to Eq.~\ref{eq:T2G}. This quantity shows a linear rise with temperature in the normal state and increases as doping is increased. The inset of Fig.~\ref{fig:T2G} plots this data as $T_{2G}^{-1}$, the spin echo decay rate. $T_{2G}^{-1}$ becomes less temperature dependent as more charge carriers are added. Otherwise, $T_{2G}$ is rather featureless in the normal state and shows no marked change upon entering the pseudogap region.

The linear increase of $T_{2G}$ depicted in Fig.~\ref{fig:T2G} is similar to data obtained on YBa$_2$Cu$_4$O$_8$ in NMR experiments reported in Fig.~3 of \cite{Corey1996} and Fig.~3 of \cite{Stern1995}, and NQR experiment (Fig.~4 of \cite{Curro1997}). The change of magnitude of a factor of $4$ from $100K$ to $700K$ is comparable in this calculation and experiment. The increase of $T_{2G}^{-1}$ as charge carriers are added is similarly observed in YBa$_2$Cu$_3$O$_{7-x}$ experiment, see e.g. Fig.~8 of Ref.~\cite{Berthier1996} and Fig.~11 of Ref.~\cite{Barzykin1995}. We find no indication of a change of slope around $\sim500K$ as discussed in Fig. 4 of Ref.~\cite{Curro1997}.

\begin{figure}[tb]
\includegraphics[width=80mm]{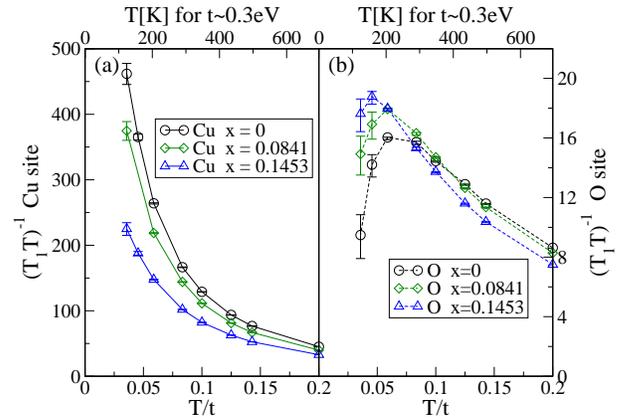}
\caption{$(T_1T)^{-1}$ plotted as a function of temperature at $U=6t$, $t'=-0.1t$, for $x=0$ to $x=0.145$, by 8-site DCA. Panel (a), solid lines: symmetry factors corresponding to $^{63}Cu$ site. Panel (b), dashed line: $^{17}O$ site (See supplemental material for explanation of uncertainties).}
\label{fig:T1_inv}
\end{figure}

Fig.~\ref{fig:T1_inv} shows the simulated spin lattice relaxation rate multiplied by the inverse temperature, $(T_1T)^{-1}$, as a function of $T$  for three dopings (see Eq.~\ref{inverseT1}) with structure factors corresponding to copper and oxygen nuclei. All results are obtained at an interaction strength of $U=6t$. $(T_1T)^{-1}$ for $^{63}Cu$ (solid line) rises rapidly when temperature is reduced. As doping is reduced, the value of $(T_1T)^{-1}$ decreases, and no clear indication of the pseudogap onset temperature is visible. In contrast, $(T_1T)^{-1}$ for $^{17}O$ (solid line) has peaks at about the same temperatures as $T^*_{Ks}$. $(T_1T)^{-1}$ for both $^{63}Cu$ and $^{17}O$ become doping indepedent at even higher temperature (see supplemental material).

While reliable results for $T_1$ from other theoretical methods are absent in the pseudogap regime, our results can directly be compared to real-frequency RPA calculations for $T_1$ in the weak coupling regime \cite{Bulut1990}. These calculations are neither limited by the momentum resolution of DCA, nor do they suffer from the limitations of analytic continuation. Therefore they provide a stringent check on the precision with which we can obtain relaxation rates. Our simulations show that $T_1^{-1}$ smoothly decreases towards zero as temperature is reduced, in good agreement with RPA $U=2t$ \cite{Bulut1990}, hinting at limitations of the random phase approximation in the intermediate coupling regime where deviations are apparent (see supplemental material).

The experimentally measured spin-lattice relaxation rates are strongly material dependent. One common feature found for the planar $Cu$ site in YBCO materials in the normal state is that $(T_1T)^{-1}$ increases slowly and linearly as $T$ decreases in a large range of temperature above  $T^*$ \cite{Takigawa1991,Tomeno1994}. As $T$ is lowered below $T^*$, it shows a decrease towards $T_c$. In contrast, experiments in LSCO materials show that $(T_1T)^{-1}$ for the planar $Cu$ site increases rapidly as temperature is decreased until $T_c$, with a larger rate as the doping level is decreased (see Ref.~\cite{Ohsugi1994}, Fig. 4). $(T_1T)^{-1}$ data for planar $^{17}O$ in LSCO are proportional to the Knight shift in the range from $100K$ to $200K$ \cite{Walstedt1994}.Doping-independent $(T_1)^{-1}$ is observed in NQR experiment on LSCO above $700K$(Fig.2 in Ref.~\cite{Slichter1993}), and NMR experiment on YBa$_2$(Cu$_{1-x}$Zn$_x$)$_4$O$_8$ above $150K$(Fig.2 in Ref.~\cite{Zheng1993}). A comparison of these two types of materials is made in Ref.~\cite{Berthier1996}. Our result is consistent with the experimental result of  LSCO and inconsistent with YBCO. We attribute this to the presence of interplanar spin couplings in the latter materials \cite{Takigawa1993}, whose existence is confirmed by neutron-scattering experiment \cite{Tranquada92}, and surmise that more complicated bilayer models might be required to yield consistent result for the YBCO spin lattice relaxation rates, also suggested from previous theoretical work\cite{millis:1993}.

In conclusion, we have shown results for the doping and temperature evolution of the knight shift, the relaxation time, and the spin echo decay time in the pseudogap regime of the two-dimensional Hubbard model. Our results were obtained using an eight-site dynamical cluster approximation calculation that treats short ranged correlations exactly and approximates longer ranged correlations in a mean field way.
These calculations show trends in temperature and doping evolution that are in remarkable agreement with experiment on single layer compounds and deviate when compare to the relaxation rate of double  layer compounds, indicating that both the relation of experimental quantities to the generalized susceptibility and the calculation of the susceptibility in the pseudogap regime are well under control.

The agreement of the calculated two-particle quantities with NMR experiment and relation to single-particle features of the pseudogap ($T^*$ and $\Delta_{pg}$)  suggests that the salient aspects of the physics of the cuprate pseudogap are contained within the simple single-orbital Hubbard model \cite{Gull10_clustercompare,Alloul14}. Phenomena absent from this calculation, {\it e.g.} stripes, multi-orbital effects or nematic order, may occur on top of the physics realized here but do not seem to be the primary cause of the pseudogap observed in the cuprates via NMR probes.

 The marked difference between multi layer and single layer materials suggests that inter-layer correlations, absent in these calculations, have a strong effect on the relaxation time. Calculating such effects is an interesting open question.

\begin{acknowledgments}
We thank R.~Walstedt and A.~J.~Millis for insightful and productive discussions on experimental data and on theory. We acknowledge the Simons Collaboration on the Many-Electron Problem and the National Science Foundation for financial support. Our simulations made use of the ALPS \cite{ALPS20} library and were performed on XSEDE using TG-DMR130036.
\end{acknowledgments}
\bibliographystyle{apsrev4-1}
\bibliography{refs.bib}
\end{document}